# Unraveling Antagonistic Collision-Controlled Reactivity in Energetic Molecular Perovskites with Deep Potential Molecular Dynamics

Ming-Yu Guo,[†] Yun-Fan Yan,[†] Pin Chen,[‡] Wei-Xiong Zhang*,[†]

[†] MOE Key Laboratory of Bioinorganic and Synthetic Chemistry, School of Chemistry, IGCME, Sun Yat-sen University, Guangzhou 510275, China.
[‡] National Supercomputer Center in Guangzhou, School of Computer Science and Engineering, Sun Yat-sen University, Guangzhou 510006, China.

**ABSTRACT:** The precise regulation of chemical decompositions in energetic materials, whether towards rapid ignition or stable endurance, requires atomic-scale principles governing reactivity, which remain elusive yet. Herein, we resolve this challenge through deep potential molecular dynamics (DPMD) simulations, uncovering a universal collision-control principle in energetic molecular perovskites, $(H_2dabco)B(ClO_4)_3$, where $H_2dabco^{2+}$ = 1,4-diazabicyclo[2.2.2]octane-1,4-diium, B = $Na^+$, $K^+$, $Rb^+$, $NH_4^+$ for DAP-1, DAP-2, DAP-3 and DAP-4, respectively. Atomic-scale simulation with Arrhenius fitting for over 100-ps trajectories reveals that increasing unreactive B-site ionic radius ($Na^+ < K^+ < Rb^+$) simultaneously reduces both activation energy $E_a$ which enhances reactivity and pre-exponential factor $A$ which suppresses collision probabilities for hydrogen transfer between site X and site A, with sharply opposing kinetic consequences. This duality well explains the peak stability and insensitivity in $K^+$-based DAP-2, which optimally balance thermal endurance and collision dissipation. For ammonium-based DAP-4, though the radius of $NH_4^+$ is close to $K^+$, the reactive B-site cation triggers hydrogen transfer that promotes C–H bond rupture. By linking static cation radii to dynamic $E_a$-ln($A$) coupling, we rationalize non-monotonic decomposition temperatures ($K^+ > NH_4^+ > Rb^+ > Na^+$) macroscopic stability and establishes cornerstones for universal energetic material design.

## INTRODUCTION

Energetic materials epitomize humanity's quest to harness chemical energy at atomic scales, having been attracting increasing attentions in recent years[1-8]. Among the emerging materials, energetic molecular perovskites with $ABX_3$-type architectures have revolutionized energy-release engineering through modular crystal engineering (**Figure 1a**). In these systems, the A-site (typically organic cations) and B-site (smaller inorganic cations) synergize with X-site oxidizing anions (e.g., $ClO_4^-$) to create integrated oxidizer-fuel frameworks.[9-12] The 1,4-**d**iazabicyclo[2.2.2]octane-1,4-diium ($H_2dabco^{2+}$) **al**kali/**a**mmonium **p**erchlorate series (DAPs), *i.e.*, $(H_2dabco)B(ClO_4)_3$, (B = $Na^+$, $K^+$, $Rb^+$, and $NH_4^+$ for DAP-1, DAP-2, DAP-3, and DAP-4, respectively) exemplifies this strategy with unparalleled tunability of detonation sensitivity and thermal stability. (**Figure 1b**)

However, rational design of such materials is impeded by significant fundamental knowledge gaps in decomposition mechanisms. On the experimental front at macroscopic level, the kinetics of initial decomposition is unachievable for regular experimental techniques, such as differential scanning calorimetry (DSC), infrared spectroscopy (IR), and thermogravimetric analysis (TGA), owing to their poor spatial-temporal resolution and precision: (i) there is almost no instrument that can characterize the rapid decomposition of energetic materials at the microscopic level and in situ, particularly since the decomposition of energetic materials itself may be destructive to delicate instruments.[13,14] (ii) Even when decomposition properties are measured, the rapid reaction kinetics introduce uncertainties, as reflected in the variability of reported data. For instance, efforts have been paid in measuring the apparent activation energy derived from the DSC exothermic peak of DAP-4, but varied by up to 20 kJ/mol across different studies,[15,16] highlighting the challenges of capturing precise kinetic parameters via instrumental techniques. At the microscopic scale, the challenge is further compounded by counterintuitive trends observed in crystallographically isostructural materials. For instance, in DAP systems, although B-site ionic radii increase monotonically following the well-established periodic trend, $Na^+ < K^+ \sim NH_4^+ < Rb^+$, which would naively suggest a corresponding monotonic trend in macroscopic properties. However, the observed behavior starkly deviates from this expectation that $K^+$-based DAP-2 exhibits peak thermal stability (364 °C) and friction sensitivity (42 N), whereas $NH_4^+$-based DAP-4 shows abrupt decomposition despite ionic radius parity with $K^+$.[12] These anomalies underscore the inadequacy of static microscopic descriptors and highlight the need for atomistic interrogation of multi-factor and dynamic coupling.

Integrating Newton's motion equation for atoms allows simulations for atomic movement, collision, and reaction, enabling successive sampling of their dynamic properties at the microscopic level, which is the principle of molecular dynamics (MD). During MD, a precise interatomic potential is essentially required. However, existing computational methodologies for MD also face common yet critical limitations. (i) Calculating the potential with density-functional theory (DFT) is inefficient, and limit the simulation size to hundreds of atoms due to the near-cubic scaling within atomic numbers in diagonalization algorithm, insufficient for capturing statistically meaningful decomposition events[17]. Although the orbital transformation (OT) algorithm[18,19] in DFT or semi-empirical quantum-chemical methods like PM6[20] bypass the diagonalization, they still fail in fundamental breakthrough in the spatiotemporal scale of the simulation. (ii) Although reactive force fields like ReaxFF[21,22] extend accessible scales, their parametrization lacks transferability especially to newly-emerging systems such as perchlorate-based perovskites.

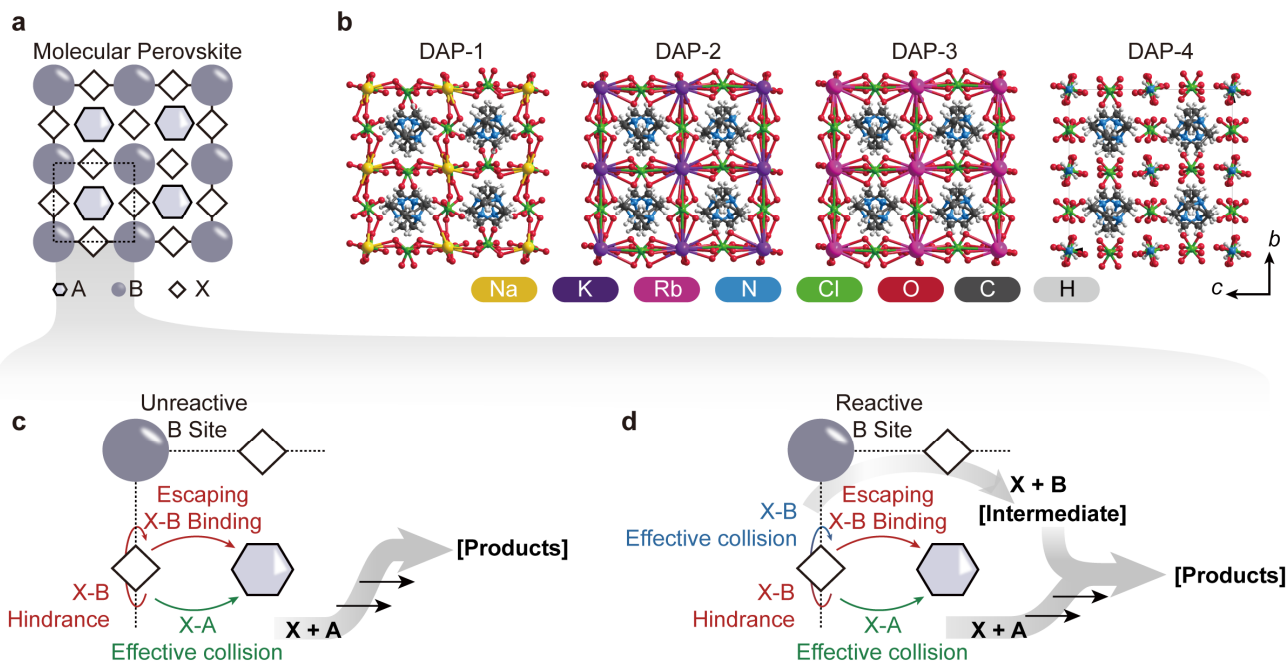

**Figure 1.** (a) The template of molecular perovskite. (b) The unit cells of DAP-1, DAP-2, DAP-3, and DAP-4. Proposed exhaustive initial inter-component decomposition paths of energetic molecular perovskites with (c) unreactive site B and reactive site B.

To address those challenges, we develop a multi-fidelity deep-learning potential (DP) with neural network achieving DFT-level accuracy with linear computational scaling, enabling thousand-atom and nanosecond-scale simulations. Such methods have been applied in chemical reactions such as fuel oxidations [23,24], heterogeneous catalytic reactions [25-27], in-solution reactions [28,29] and decompositions in organic solid energetics [30-33], which on one hand justify the feasibility of such methods; on the other hand, they lack systematic comparison and conceptualizations. To fill the gaps, we selected the DAP-1 to DAP-4 series as a set of highly isostructural materials for mechanistic investigation, allowing us to rigorously control variables and isolate the effects of B-site cations on decomposition behavior. By systematically enumerating all possible initial inter-component reaction pathways in the DAP series, we identified two distinct scenarios: (i) pathways dominated by X–A interactions when the site B remains unreactive (**Figure 1c**), and (ii) pathways involving active participation of the B-site, which can significantly alter the decomposition kinetics (**Figure 1d**). By focusing on these dominant pathways and coupling DP-driven trajectory analysis with *ab initio* electronic structure analysis, our work uncovered the fundamental principles governing the decomposition of energetic molecular perovskites, revealing that the interplay of two opposing yet complementary kinetic factors: X–A attraction and B-site steric effects, dictates the decomposition behavior. These insights not only resolve the counterintuitive trends observed across the DAP series but also provide a significant clue for the rational design of next-generation energetic materials.

## RESULTS AND DISCUSSION

*Development and Evaluation of DP Models*

To address the aforementioned efficiency-accuracy dilemma in studying the decomposition mechanisms, we developed a robust deep-learning potential (DP) model using an active learning framework. This model encompasses 20 energetic materials, including the DAP series (all materials are listed in **Figure S1**), ensuring a broad and representative sampling space.

Deep learning is recognized as a universal function approximator fitting intricate data relations [34], forms the foundation of our approach. In deep potential molecular dynamics (DPMD), the aim is to accurately fit DFT-level potential energy surface defined by energy $E$, force $F$, virial $V$ with respect to structure defined by lattice $L$, atom types $A$, and atomic coordinates $X$ for reliable MD simulations [35,36]. (**Figure 2a**) For model training, the quality of dataset is crucial. First, it should cover the sampling space, achievable through active learning,[37-39] (**Figure 2b**) which enables sampling diverse-distributed datasets with minimal volume. Our iterations of active learning were conducted in a progressive strategy (**Table S1**) of potentials effective up to 4000 K and 120 GPa, resulting in 87020 frames of reasonable-distributed spin-polarized DFT labels (**Figure S5**) with overall accuracy up to 98.3%. (see **Figure S6** for per-system deviation distribution) Second, the self-consistency should be ensured, which we achieved by rigorous convergence test for DFT calculations. (**Figure S2–S4**)

A high-quality DP model requires (i) high accuracy in directly predicting DFT labels, (ii) high efficiencies running inferences, and (iii) conservation in potential energy surface.

To meet these requirements, we employed the DeepMD architecture[40-42], which prioritizes energy conservation and ensures the stability of long MD simulations. Before formal MD running, we conducted extensive evaluations to validate its performance. Our DP model demonstrates ca. 100 times faster than DFT-MD and superior to semi-empirical PM6. (**Figure 2c**) Direct label predictions on the test set yielded root mean square errors (RMSE) of 26.6 meV/atom for energy, 375.7 eV/Å for force, and 21.7 meV/atom for the virial coefficient (**Figure 2d**), better than chemical accuracy energy-wise (1 kcal/mol). Furthermore, the conservativeness of the yielded DP models was confirmed by evaluating energy drifts and standard deviations in *NVE*-MD, thereby validating its MD simulation reliability (**Figure S7**).

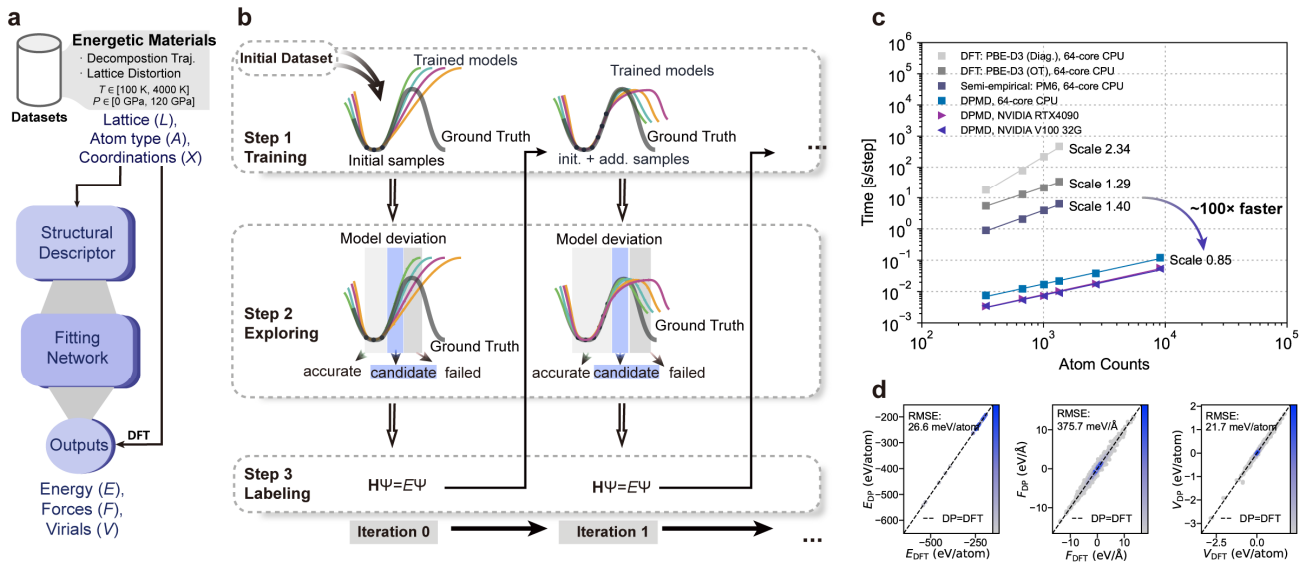

**Figure 2.** (a) The schematic illustration of DP model. (b) The flowchart of active learning-based dataset sampling. The model evaluations involving (c) efficiency evaluations where CPU tests were conducted on Intel Xeon Platinum 8358P CPU @ 2.60GHz; GPU tests were conducted on NVIDIA RTX 4090, 24GB and NVIDIA V100, 32GB; (d) direct inference on test set results compared to DFT where the color mapping represents the kernel density estimation.

*Species and Reaction Analysis*

To elucidate the decomposition mechanisms of the DAP series, we conducted molecular dynamics (MD) simulations in the temperature range of 1666 to 2500 K using 3×3×3 supercells of DAP-1 to DAP-4. The simulations were performed in the *NVT* ensemble after energy minimization and *NPT* equilibration at 300 K and 1 bar. To monitor the precision of the MD process on-the-fly, we calculated the average deviations for forces, which were consistently less than 5% (**Figure S8**) This rigorous validation ensures the reliability of our simulation results and provides a robust foundation for mechanistic analysis.

Accurate identification and tracking of chemical species during decomposition are critical for understanding the reaction mechanisms. To achieve this, two distinct methods were employed to track species evolution during decomposition. The first one is that for simple inorganic substances, it can be directly defined by the coordination number and cutoff radius outside the central atom (criteria see **Table S2**). For complex organic substances, it is necessary to further search using a depth-first search algorithm, which is implemented by calling the OpenBabel library[43] in ReacNetGenerator by Zeng et al[44] (criteria see **Table S3**) and recorded in SMILES encoding.

We define "the completion of initial reactions" as the reaction between site X and site A, *i.e.*, in this scenario, perchlorate and H$_2$dabco is completed, so that the statistical number tends to 0. (**Figure S9–S12**) Through the above statistical method, we found that at 1666 K, perchlorate (X) and H$_2$dabco (A) can be completely reacted within 100 ps. Furthermore, by tracking the species evolution of each atom in different frames, the reaction path can be constructed.[44]

First, we investigate the dominant decomposition pathway of H$_2$dabco, focusing on their behavior within various DAP variants. Our statistical analysis reveals that the primary reactions involve two inter-component hydrogen transfer reactions and one intra-component ring-opening reaction: (i) deprotonation of N–H and hydrogen transfer between perchlorate and the H$_2$dabco, (ii) bond homolysis of C–H and hydrogen transfer between perchlorate and H$_2$dabco, and (iii) cleavage of the C–N bond and ring opening. Consistently, across all four DAP variants examined, the hierarchy of these pathways is maintained, with N–H deprotonation emerging as the most prominent mechanism, followed by C–H homolysis, and finally C–N ring opening (**Figure 3a–d**). At relatively low simulation temperatures (e.g., 1666 K), more than 60% of the H$_2$dabco in all four materials undergoes N–H deprotonation, about 30% undergoes C–H homolysis, and C–N ring opening accounts for at most 5%. As the temperature increases, the distribution of products shifts significantly, with the reaction ratios for C–H cleavage and C–N bond cleavage increasing. For instance, at 2500 K, the proportion of C–N ring opening rises to approximately 20%. Nevertheless, the reactions remain predominantly governed by N–H and C–H cleavage.

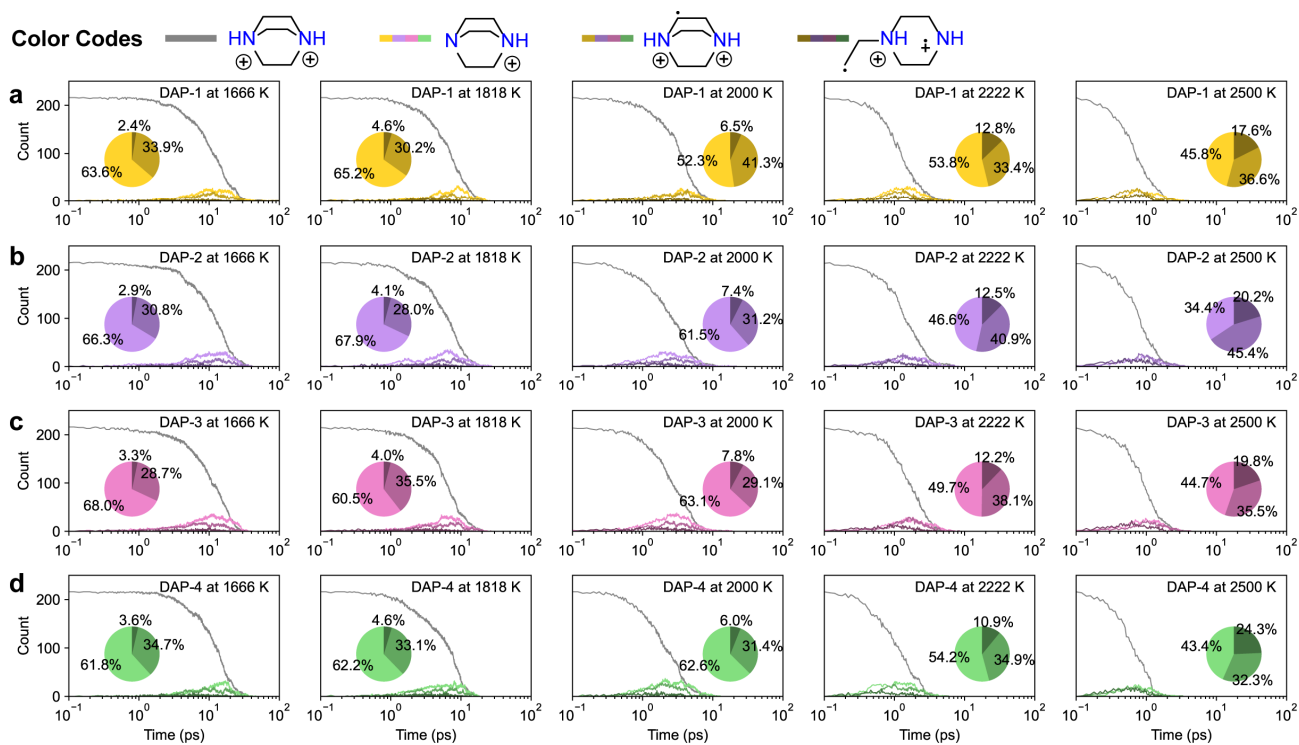

**Figure 3.** Key decomposition species from H₂dabco at various temperatures in (a) DAP-1, (b) DAP-2, (c) DAP-3, and (d) DAP-4. Insets are cumulated species counts.

To gain deeper insights, we employed the protocol by Wu et al. [45] to calculate the reaction rates of the chemical reactions corresponding to these three pathways at different temperatures and performed Arrhenius fitting. (**Figure 4a**) The results are consistent with chemical intuition: the activation energy of the N–H bond is in the range of 55.69 ± 5.40 to 64.10 ± 8.17 kJ/mol (**Figure 4b**), because N–H deprotonation is an acid-base reaction; the activation energy of the C–H bond cleavage is the highest, ranging from 109.64 ± 2.69 to 120.29 ± 6.50 (**Figure 4c**); so the latter is sensitive to temperature and its proportion increases at high temperatures. The activation energy of the C–N bond cleavage is in the range from 68.64 ± 2.98 to 72.11 kJ/mol (**Figure 4d**), but the activation entropy of the C–N cleavage is greater than the collision reaction of the bimolecular reaction due to the release of degrees of freedom and spin, so its proportion also increases with increasing temperature.

Next, we study the regularity of the kinetic parameters of different materials in each elementary reaction. Comparing DAP-1, DAP-2, and DAP-3, we find that with the increase of the radius of the cation at the B site, the activation energy and pre-exponential factor of the C–H homolysis and N–H deprotonation reactions decrease respectively; comparing DAP-2 and DAP-4, we find that the activation energy and pre-exponential factor are close. (**Figure 4e–f**) For ring opening on C–N, the difference is relatively not prominent. (**Figure 4g**) This means that the reaction kinetic parameters at the X and A sites are highly correlated with the properties of the B site, mainly the radius. However, these two results have opposite effects on the X–A reaction. First, the activation energy is reduced, which means that X–A only needs to cross a lower energy barrier to react, but the reduction in the pre-exponential factor means that the collision probability is not high. In other words, although the monotonic B-site properties point to two monotonic changes in kinetic parameters, the effects of these two kinetic parameters on the apparent reactivity are opposite. It is likely that this leads to the optimal balance of macroscopic thermal decomposition temperature and friction sensitivity. The relatively minor differences in the activation energy and pre-exponential factor for C–N ring opening across DAP-1, DAP-2, and DAP-3 (Figure 4g) can be attributed to the intrinsic nature of the reaction and the indirect influence of the B-site cation.

Unlike N–H deprotonation and C–H homolysis, which are more sensitive to the electronic and steric effects of the B-site cation, C–N ring opening is primarily an intramolecular process involving the cleavage of the C–N bond within the H2dabco ring. This reaction is less dependent on the interactions between the site X (perchlorate) and site A (H₂dabco), and thus the properties of the B-site cation (e.g., radius, charge density) have a weaker influence on its kinetics. Additionally, the C–N bond cleavage involves the release of degrees of freedom and spin, which contributes significantly to the activation entropy but is less affected by the external environment created by the B-site cation. As a result, the differences in the kinetic parameters for C–N ring opening are less pronounced compared to those for N–H and C–H reactions.

Complementary to these kinetic analyses, representative MD trajectory snapshots captured at 2000 K (**Figure 4h–j**) visually depict the hydrogen transferring processes on the N–H and C–H bonds as well as the C–N ring opening events at the atomic level. These snapshots clearly highlight the key reactive events, with specific particle identifiers marking the species involved, thereby reinforcing the proposed reaction pathways.

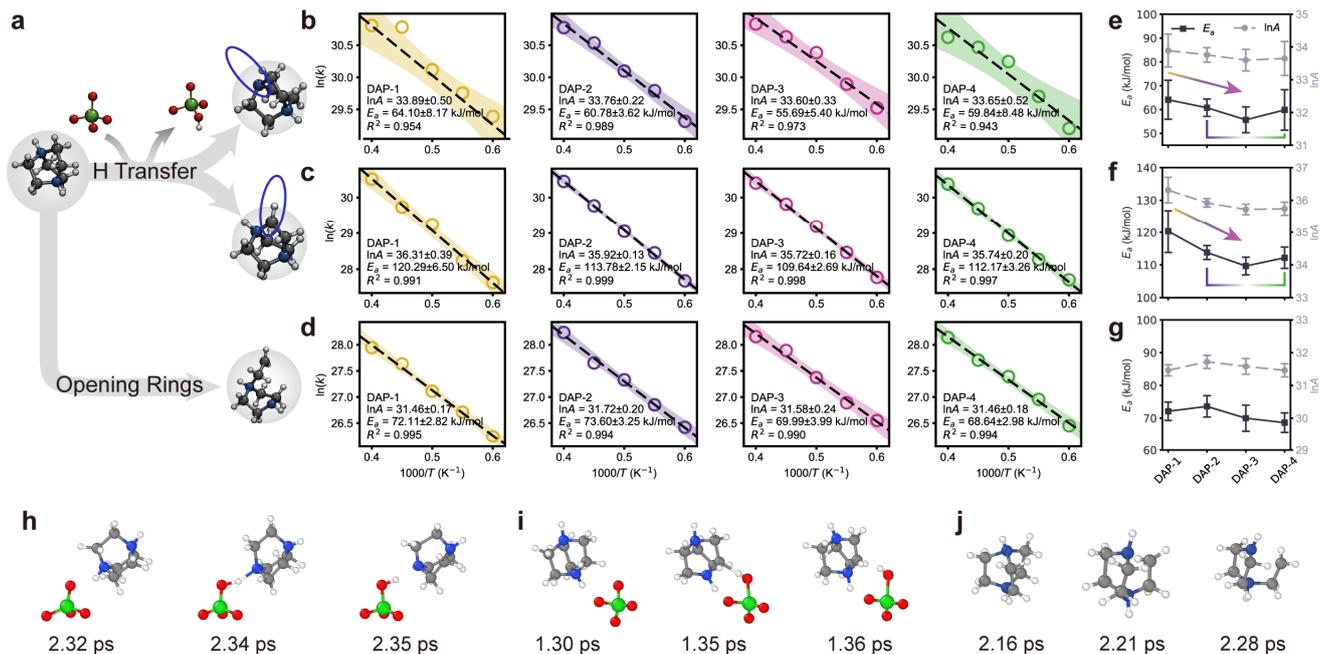

**Figure 4.** (a) Schematic illustration of hydrogen transfer pathway of H$_2$dabco. The Arrhenius fitting of hydrogen transfers on (b) N–H and (c) C–H, where the shades denote the 95% confident interval. (d–e) The relevant fitted activated energies and pre-exponential factors for different materials. Comparison of fitted activation energies and pre-exponential factors for DAPs for (e) hydrogen transfer involving N–H, (f) hydrogen transfer involving C–H, and (g) ring opening on C–N; The representative MD snapshots from DAP-4 at 2000 K of (h) deprotonation on N–H (The Particle Identifiers for N and O are 5585 and 8474, respectively), (i) homolysis on C–H (The Particle Identifiers for C and O are 4138 and 8784, respectively), and (j) ring opening on C–N (The Particle Identifiers for C and N are 4002 and 5298, respectively).

*Influence of Static Properties of Site B*

To further analyze those trends, we examined the non-chemical kinetic effects of B-site radius. Since the site B is non-reactive in DAP-1, DAP-2, and DAP-3, "ineffective collisions" between the X site and B site compete with the effective collisions of X–A. We defined the collision frequency (C.F.) of X–B as follows:

$$\text{C. F.} = \frac{C_{B,coll}}{N_X * T_{sim}} \quad (1)$$

where $C_{B,coll}$ represents the counts of collision events for B, defined as those events where the distance between the X and B sites is less than a threshold determined from the ionic radii (set to 1.2 times $r_B$ as listed in **Table S4**); $N_X$ represents the counts of site X, placed in denominator for normalization; $T_{sim}$ represents the simulation time span for statistics. As shown in **Figure 5a**, C.F. in different time intervals at 1666 K shows that in the early stage of the reaction (simulation time < 0.1 ps), the collision frequency exhibits a pronounced positive correlation with the ionic radius. The observed trend was DAP-3 > DAP-4 ~ DAP-2 > DAP-1, suggesting that larger B-site cations increase the likelihood of ineffective collisions. This, in turn, thereby reduces the probability of the more productive X–A collisions that drive the reaction forward.

From the perspective of activation energy, the effect of the B site on the X–A reaction appears to be predominantly mediated by the attractive interactions between the X and B sites. To investigate this phenomenon, we employed electron density basin analysis[46-48] to characterize the spatial region occupied by the B-site cations. **Figure 5b** displays the electron density isosurfaces (isovalue = 1.0 a.u.) in the basin for Na$^+$, K$^+$, Rb$^+$ and NH$_4^+$, showing the extent of the electron density basins correlates with the ionic radii in the order, which is Na$^+$ < K$^+$ ~ NH$_4^+$ < Rb$^+$. When the geometric center separation between the X site and the B site is minimal, a larger atomic volume for the B-site cation leads to an enhanced interaction with the electron cloud associated with the X site. This interaction is predominantly governed by electrostatic attraction and inductive forces, with minimal contributions from dispersion effects, and is further modulated by mutual repulsion due to orbital exchange.

To further validate those observations, energy decomposition analysis was performed using symmetry-adapted perturbed theory (SAPT).[49,50] The results revealed that while the three attractive forces (electrostatic, inductive, and dispersion) between X and B are relatively similar across the four materials, with variations of less than 6 kJ/mol. The most significant divergence arises from exchange repulsion, which follows the order of Na$^+$ < K$^+$ < NH$_4^+$ < Rb$^+$. Consequently, the overall binding energy, when interacting with perchlorate, adheres to the trend Na$^+$ > K$^+$ > NH$_4^+$ > Rb$^+$ (**Figure 5c**). This observation is in strong agreement with the experimentally determined activation energies, thereby substantiating the hypothesis that the attractive X–B interaction acts as an impediment to the progress of the effective X–A reaction.

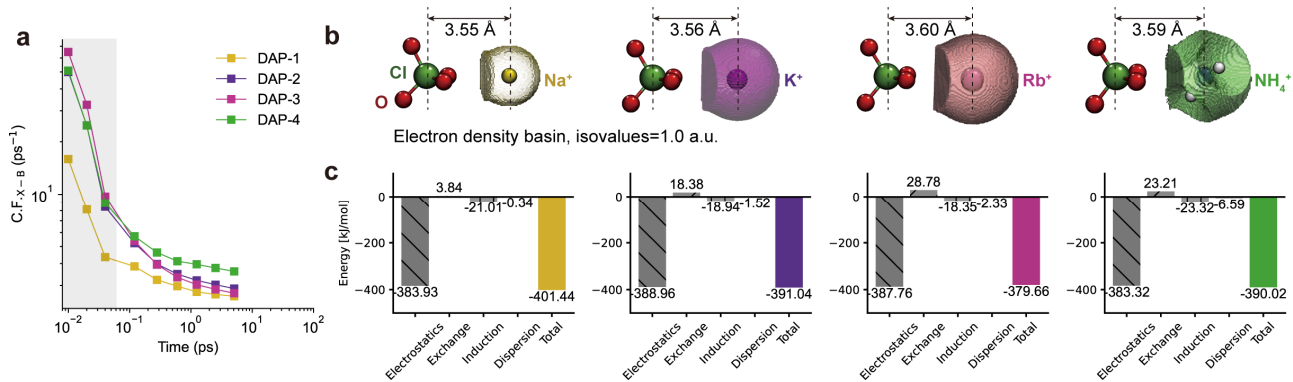

**Figure 5.** Interaction analysis of site X and site B. (a) Collision frequency (C.F.); (b) Basin analysis of electron density (isovalue = 1.0 a.u.); (c) SAPT decomposition results.

*Influence of Reactivity of Site B*

The reactivity of the B-site cation plays a pivotal role in dictating the decomposition pathways and kinetics of energetic molecular perovskites. By comparing DAP-2 and DAP-4, we observe that although the ammonium ion's radius is similar to $K^+$, its reactivity as a hydrogen carrier significantly modifies the reaction pathway. In DAP-4, ammonia, formed through the deprotonation of the B-site ammonium ion, actively participates in the decomposition process, leading to a dramatic reduction in the activation energy of key reactions. Specifically, the activation energy for C–H bond cleavage with ammonia in DAP-4 is reduced to 51.76 ± 8.66 kJ/mol, which is almost half of that with the participation of perchlorate. This substantial decrease underscores the catalytic role of ammonia in facilitating hydrogen transfer and lowering the energy barrier for bond dissociation.

Further statistical analysis of ammonium ions and ammonia dynamics (**Figure S12**) provides additional insights into the reaction mechanism. During the early stages of decomposition, the concentration of ammonium ions decreases, while ammonia initially increases. This trend reverses in the later stages, with ammonium ions increasing and ammonia exhibiting a fluctuating pattern. Notably, the point at which ammonium ion concentration reaches its minimum coincides precisely with the peak in $H_2$dabco C–H homolysis products. This correlation suggests that $H_2$dabco acts as a hydrogen donor in the ammonia-ammonium equilibrium, consuming ammonia and driving the equilibrium toward ammonium consumption.

This dual role of $H_2$dabco, both as a reactant and a hydrogen donor, underscores the complexity of the decomposition process in DAP-4. The ability of ammonia to participate in hydrogen transfer reactions not only accelerates bond cleavage but also introduces a dynamic equilibrium that influences the overall reaction kinetics. These findings further demonstrate that the reactivity of the B-site cation is not merely an intrinsic static property; rather, it is a dynamic factor that can profoundly reshape both the decomposition pathway and the associated energetics.

## CONCLUSION

In this study, we have employed deep potential molecular dynamics (DPMD) simulations to unravel the intricate collision-controlled reactivity in energetic molecular perovskites, specifically the $(H_2dabco)B(ClO_4)_3$ series (DAP-1 to DAP-4). Our findings reveal a universal yet antagonistic kinetic principle that governs the decomposition behavior of these materials, bridging the gap between microscopic dynamics and macroscopic properties. By systematically analyzing the interplay between B-site ionic radii, activation energy ($E_a$), and pre-exponential factor ($A$), we have elucidated the non-monotonic thermal stability and sensitivity trends observed in the DAP series.

The key insight lies in the dual role of the B-site ionic radius: increasing ionic radius ($Na^+ < K^+ < Rb^+$) simultaneously reduces the activation energy, which enhances reactivity, and decreases the pre-exponential factor, which suppresses collision probabilities. This antagonistic coupling explains the peak thermal stability and insensitivity of $K^+$-based DAP-2, which optimally balances these opposing effects. In contrast, the ammonium-based DAP-4, despite its ionic radius parity with $K^+$, exhibits distinct behavior due to the reactive nature of $NH_4^+$, which facilitates hydrogen transfer and promotes C–H bond rupture, leading to abrupt decomposition.

Our results not only resolve the counterintuitive trends in the DAP series but also establish a robust framework for understanding the dynamic coupling between static structural descriptors and kinetic parameters. By linking microscopic atomic interactions to macroscopic material properties, this work provides a cornerstone for the rational design of next-generation energetic materials with tailored thermal stability and reactivity. Future research could focus on extending this framework to more complex energetic material systems and exploring the impact of external stimuli on the decomposition mechanisms.

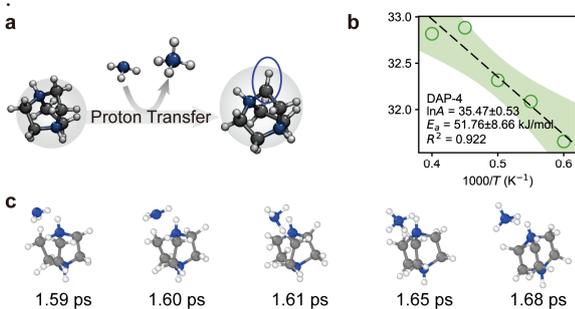

**Figure 6.** (a) Schematic illustration of hydrogen transfer pathway of $H_2$dabco interacted with ammonia of DAP-4. (b) the Arrhenius fitting of hydrogen transfer on C–H, where the shades denote the 95% confident interval. (c) The representative MD snapshots for DAP-4 at 2000 K (The Particle Identifiers for C and N are 4427 and 5745, respectively).

## ASSOCIATED CONTENTS

*Supporting Information*

Additional details of theoretical and experimental results are recorded in Supporting Information.

*Data Accessibility and Reproducibility*

The datasets for DP training and trained DP models have been uploaded to AIS Square:

**Datasets:** https://aissquare.com/datasets/detail?pageType=datasets&name=EnergeticMaterials-v1&id=311

**Models:** https://aissquare.com/models/detail?pageType=models&name=DAP1-L0_EnergeticMaterials-v1&id=312.

Codes are openly available in a public repository:

**Codes:** https://github.com/SchrodingersCattt/DPA-EMs

Other major datafiles are available on request from the authors.

*AUTHOR INFORMATION*
**Corresponding Author**
*Wei-Xiong Zhang − MOE Key Laboratory of Bioinorganic and Synthetic Chemistry, School of Chemistry, IGCME, Sun Yat-sen University, Guangzhou 510275, China; Orcid https://orcid.org/0000-0003-0797-3465; Email: zhangwx6@mail.sysu.edu.cn;

**Authors**
**Ming-Yu Guo** − MOE Key Laboratory of Bioinorganic and Synthetic Chemistry, School of Chemistry, IGCME, Sun Yat-sen University, Guangzhou 510275, China; Orcid https://orcid.org/0009-0008-3744-1543;
**Yun-Fan Yan** − MOE Key Laboratory of Bioinorganic and Synthetic Chemistry, School of Chemistry, IGCME, Sun Yat-sen University, Guangzhou 510275, China;
**Pin Chen** − National Supercomputer Center in Guangzhou, School of Computer Science and Engineering, Sun Yat-sen University, Guangzhou 510006, China.

**Notes**
The authors declare no competing financial interest. DeepSeek-v3[51] was utilized for language polishing; all artificial intelligence-generated contents have been finally checked by human authors.

*ACHKNOWLEGEMENT*

This study was funded by National Natural Science Foundation of China (U2341287 and 22488101), Guangzhou Science and Technology Program (2024A04J6499), and Fundamental Research Funds for the Central Universities, Sun Yat-Sen University (23lgzy001). The funder played no role in study design, data collection, analysis and interpretation of data, or the writing of this manuscript. The computational resource was supported by the Bohrium Cloud Platform (https://bohrium.dp.tech/) and National Supercomputing Center in Guangzhou (NSCC-GZ, Tianhe-2).

**TABLE OF CONTENTS**

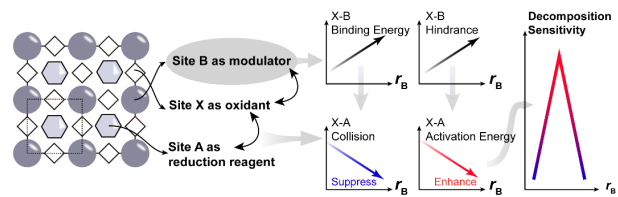

# Unraveling Antagonistic Collision-Controlled Reactivity in Energetic Molecular Perovskites with Deep Potential Molecular Dynamics

Ming-Yu Guo,[†] Yun-Fan Yan,[†] Pin Chen,[‡] Wei-Xiong Zhang*,[†]

[†] MOE Key Laboratory of Bioinorganic and Synthetic Chemistry, School of Chemistry, IGCME, Sun Yat-sen University, Guangzhou 510275, China.
[‡] National Supercomputer Center in Guangzhou, School of Computer Science and Engineering, Sun Yat-sen University, Guangzhou 510006, China.

# I. Methods

**Preparation for Datasets.**

The crystal structures are collected from Cambridge Crystal Data Center (CCDC) or appendixes in documents (**Figure S1**).

The QuickStep module of the CP2K package [1] is used for all the DFT calculations. Since the quality of DFT data is crucial., our research started with a series of convergence test. Considering the balance of computational cost, we utilized proper supercell to allow gamma point-only calculations with orbital transformation (OT) [2,3] approach for DFT calculations. So the convergence test was mainly conducted for cutoff energies (**Figure S2–Figure S4**) and finalize the parameter to be 800 Ry with relative cutoff 60 Ry. During DFT calculations, unrestricted shell was considered to appropriately describe generated radicals with explicit spins. All atoms are described using the MOLOPT basis set in combination with norm-conserving Goedecker-Teter-Hutter (GTH) pseudopotentials. The Perdew-Burke-Ernzerhof (PBE) [4] exchange-correlation functional is employed throughout and the electron density is written over an auxiliary plane-wave basis set with appropriate energy cut-off. The D3 dispersion correction was employed to appropriately correct the inter-molecular non-covalent interactions[5,6]. The self-consistent field (SCF) convergence of both the outer and inner loops is achieved with an accuracy of $10^{-5}$ Hartree.

Initial dataset was built with short DFT-MD simulations for each system, at pressure of $10^5$ bar and temperature of 300 K and 2000 K under the *NPT-F* ensemble and 3500 K under *NVT* ensemble with time step as 0.1 fs for 500 fs. The structure (i.e. the lattices, atomic types and coordinates) and the relevant labels (i.e. the energies, forces, and virials) were properly recorded. To avoid redundancy, the initial datasets are selected based on temperature as follows: for trajectories at 300 K, the ratios was 1%; the others were 5%.

**Active Learning.**

Then the active learning process was triggered. In each iteration of the active learning, four different DP models were trained with different random seeds for initial weights of the neural networks. After this, a short MD simulation was performed based on the first DP model, and the potential energies and atomic forces were also predicted by the other three models. For each atom in a given snapshot, we can calculate the force relative model deviation by

$$R_i = \frac{\varepsilon}{\langle f_i \rangle + d}, \quad \overline{f}_i = \langle f_i \rangle \qquad (S1)$$

where $\langle f_i \rangle$ denotes the average forces on the atom *i* of the four DP models, $\varepsilon$ denotes the maximal standard deviation of the atomic force. $d$ is a constant used to make sure that the atom with a small $\varepsilon$ will also have a small relative model deviation, and it was set to 1.0 eV/Å in this study. Because atomic forces in thermal decomposition have a large range of values, the relative model deviation instead of the absolute one was employed for selecting structures during the simulation. If $R_i$ is less than the lower trust level, the snapshot will be labeled as "accurate", which means that it has been accurately learned by the neural network; while if the $R_i$ of a given snapshot is within a certain range (lower trust level < $R_i$ < higher trust level), this snapshot will be labeled as "candidate"; else the snapshot was labeled as "fail" due to the potential crashed simulation. In each iteration, we randomly selected at most 250 frames of "candidate" structures. The potential energy, atomic forces and virials of those selected frames were calculated by DFT and added back to the dataset. In each iteration, we appropriately increased the length of the simulation as listed in **Table S1**.

**Training.**

In this study, we employed the DeepPot-Attention architecture (DPA-1) [7] without the attention layer (noted as DPA-1-L0), implemented in the DeePMD-kit code package (version 2.2.9) [8-10]. In this method, the potential energy $E$ of each atomic configuration of the system is expressed as the sum of "atomic energies" $E = \sum_i E_i$, where $E_i$ depends on the chemical environment of the atom $i$, which is defined by the positions of all the neighbors of the atom $i$ in a sphere with a cut off radius $R_c$.

The neural network is used to construct the functional relationship between the chemical environment and the atomic energy. For both systems, the NN model includes an embedding network and a fitting network, among which the fitting net used the ResNet architecture. The embedding network was designed to preserve the translational, rotational, and permutational symmetries of the system, and the fitting network maps the embedded features to the atomic energy. In this work, the size of the embedding network was set to (25, 50, 100) and the number of columns of the submatrix of the embedding matrix was set to 12, while the size of the fitting network was set to (240, 240, 240). The cutoff radius was set to 6.0 Å and the descriptors decay smoothly from 0.5 Å to 6.0 Å. The training step was 4000000, with the initial learning rate set to 0.001 and decaying every 20000 steps to finally $3.51 \times 10^{-8}$. The loss is defined by

$$L(p_e, p_f, p_v) = \frac{p_e}{N}\Delta E^2 + \frac{p_f}{3N}\sum_i \Delta \mathbf{F}^2 + \frac{p_v}{9N}\|\Delta \mathbf{V}\|^2 \quad (S2)$$

where the $N$ denotes the atom number, $\Delta$ denotes the root-mean-squared error (RMSE) between predicted label and ground truth, $E$ denotes energy per atom, $F$ denotes force per atom on each direction, and $\mathbf{V}$ denotes the virial tensor. Coefficients $p_e$, $p_f$, and $p_v$ are tunable pre-factors, adjusted from 0.1, 1000, and 0.2 to 1.0, respectively. Model compression[11] was utilized to accelerate inference procedure.

**Molecular Dynamics.**

MD studies were conducted on the Large-scale Atomic/Molecular Massively Parallel Simulator (LAMMPS) code package[12,13]. Thermal decomposition studies were performed in the NVT ensemble with a Nosé-Hoover thermostat at target temperatures. Energy minimization was conducted with specific criteria, followed by 1 ps of *NPT* pre-equilibrium simulations with an isotropic barostat at 1.0 bar and 300 K. The time step was set to 0.1 fs, with temperature damping parameter was set to 100 (0.01 ps) for both *NPT* and *NVT* simulation, while pressure damping parameter was set to 1000 times the time step (0.1 ps), respectively.

**Quantum Chemical Calculation.**

(i) The spin population calculations were calculated with Gaussian 16[14] with explicitly setting unrestricted PBE0-D3[4-6,15] functional with def2-TZVP[16] basis sets (noted as PBE0-D3/def2-TZVP).

(ii) The basin analysis for eletron density was calculated with Multiwfn 3.8dev [17,18] with wavefunctions obtained from Gaussian 16, level PBE0-D3/def2-TZVP.

(iii) Zeroth-order symmetry-adapted perturbation based on density fitting (DF-SAPT0) calculations[19,20] were conducted with PSI4[21] with basis set def-TZVP in considerations of support with heavy atoms like Rb.

**Visualization Softwares and Python Packages.** Diamond 4.6.8[22], VMD 1.9.3[23], Ovito 3.9.1[24]; Python 3.11, TensorFlow 2.9.0 [25], deepmd-kit 2.2.9 [8,9], dpgen 0.12.0 [26], reacnetgenerator 1.6.15 [27], cp2kdata 0.7.1, dpdata 0.2.21, MDAnalysis 2.8.0 [28-30], PSI4 1.9.1[21], scipy 1.13.1 [31], pymatgen 2024.6.10 [32], ase 3.24.0 [33], matplotlib 3.9.0 [34].

# I. Supplementary Figures

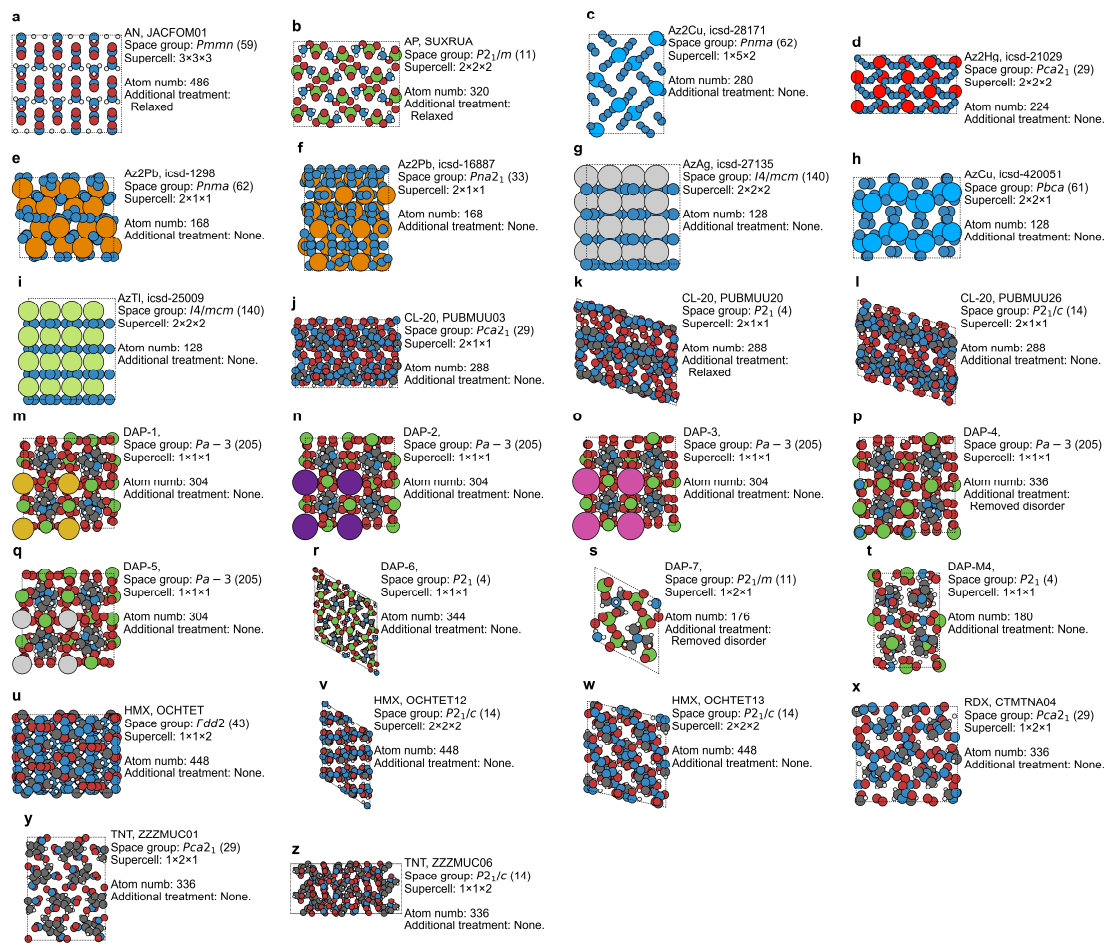

**Figure S1.** The dataset-covered 20 energetic materials within 26 structures due to polymorphs. (**a**) AN: ammonium nitrate; (**b**) AP: ammonium perchlorate; (**c**) Az2Cu: copper(II) azide; (**d**) Az2Hg: mercury(II) azide; (**e–f**) Az2Pb: lead (II) azide; (**g**) AzAg: silver(I) azide; (**h**) AzCu: copper(I) azide; (**i**) AzTl: thallium(I) azide; (**j–l**) CL-20; (**m**) DAP-1: $(H_2dabco)Na(ClO_4)_3$[35]; (**n**) DAP-2: $(H_2dabco)K(ClO_4)_3$[35]; (**o**) DAP-3: $(H_2dabco)Rb(ClO_4)_3$[35]; (**p**) DAP-4: $(H_2dabco)(NH_4)(ClO_4)_3$[35]; (**q**) DAP-5: $(H_2dabco)Ag(ClO_4)_3$[36]; (**r**) DAP-6: $(H_2dabco)(NH_3OH)(ClO_4)_3$[37]; (**s**) DAP-7: $(H_2dabco)(NH_3NH_2)(ClO_4)_3$[37]; (**t**) DAP-M4[38]: $(MeHdabco)(NH_4)(ClO_4)_3$; (**u–w**) HMX: 1,3,5,7-Tetranitro-1,3,5,7-tetrazocane (a.k.a. octogen); (**x**) RDX: 1,3,5-Trinitroperhydro-1,3,5-triazine (a.k.a hexogen); (**y–z**) TNT: 2-methyl-1,3,5-trinitrobenzene.

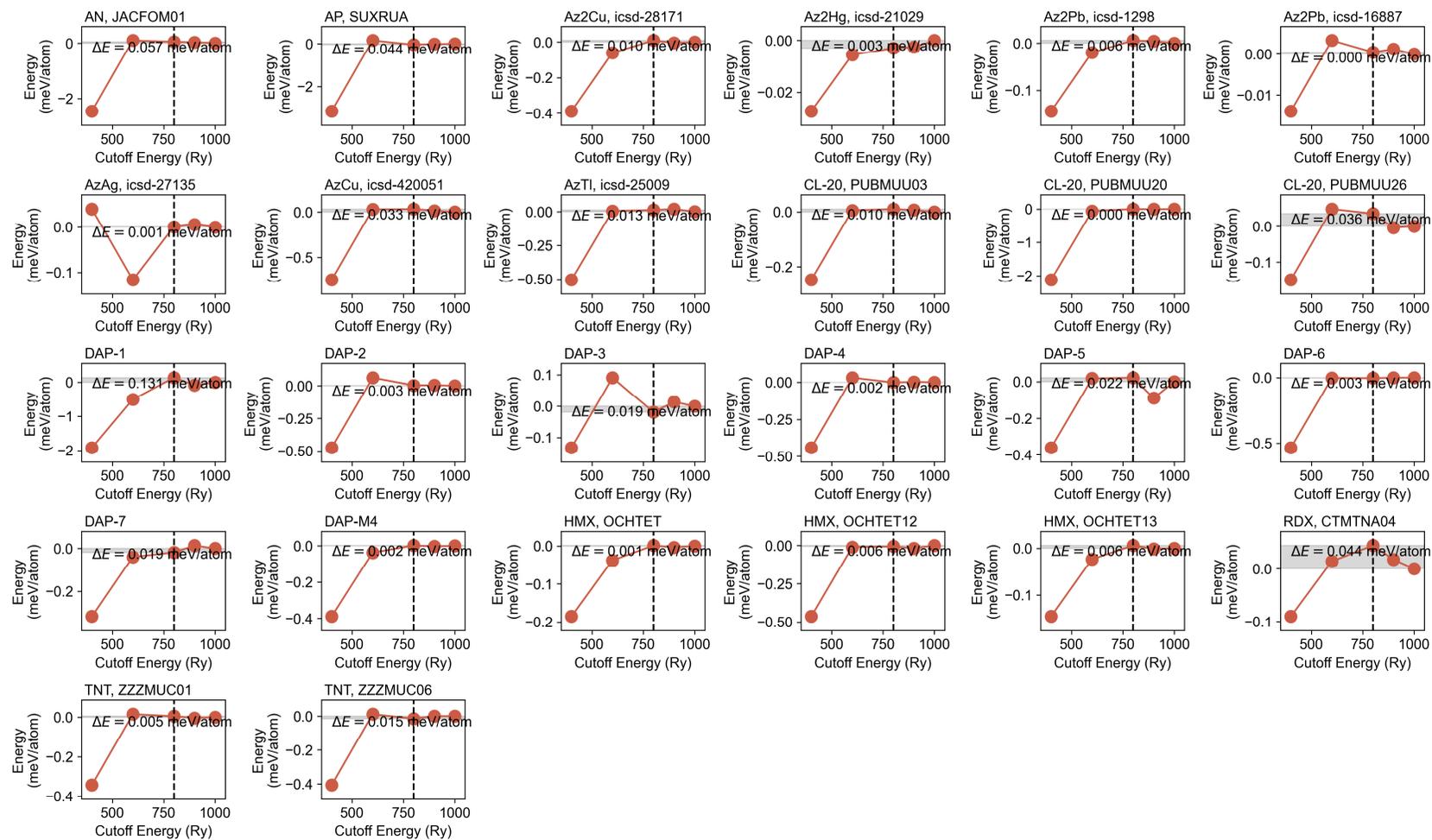

**Figure S2.** The convergence test with respect to cutoff energy of energies.

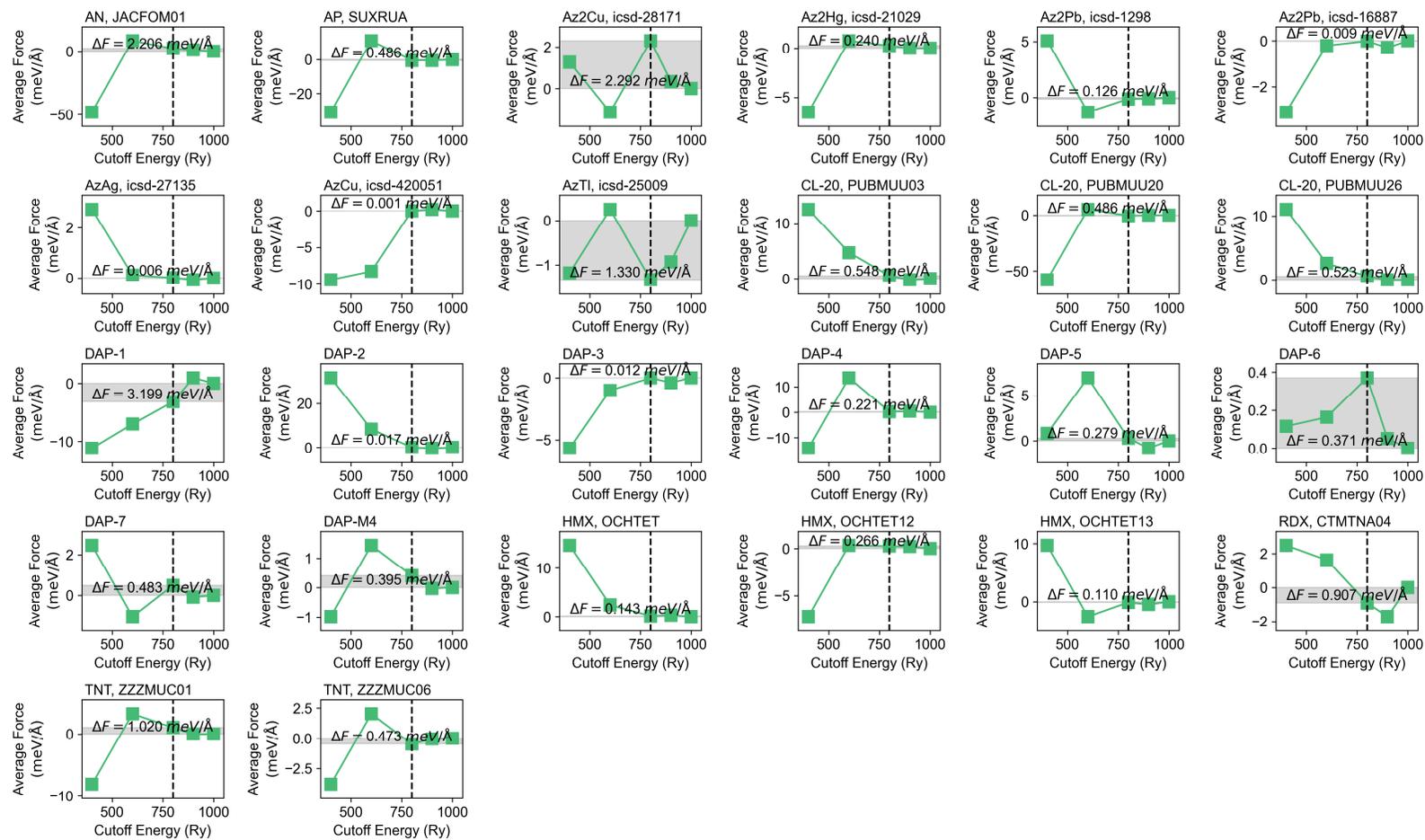

**Figure S3.** The convergence test with respect to cutoff energy of forces.

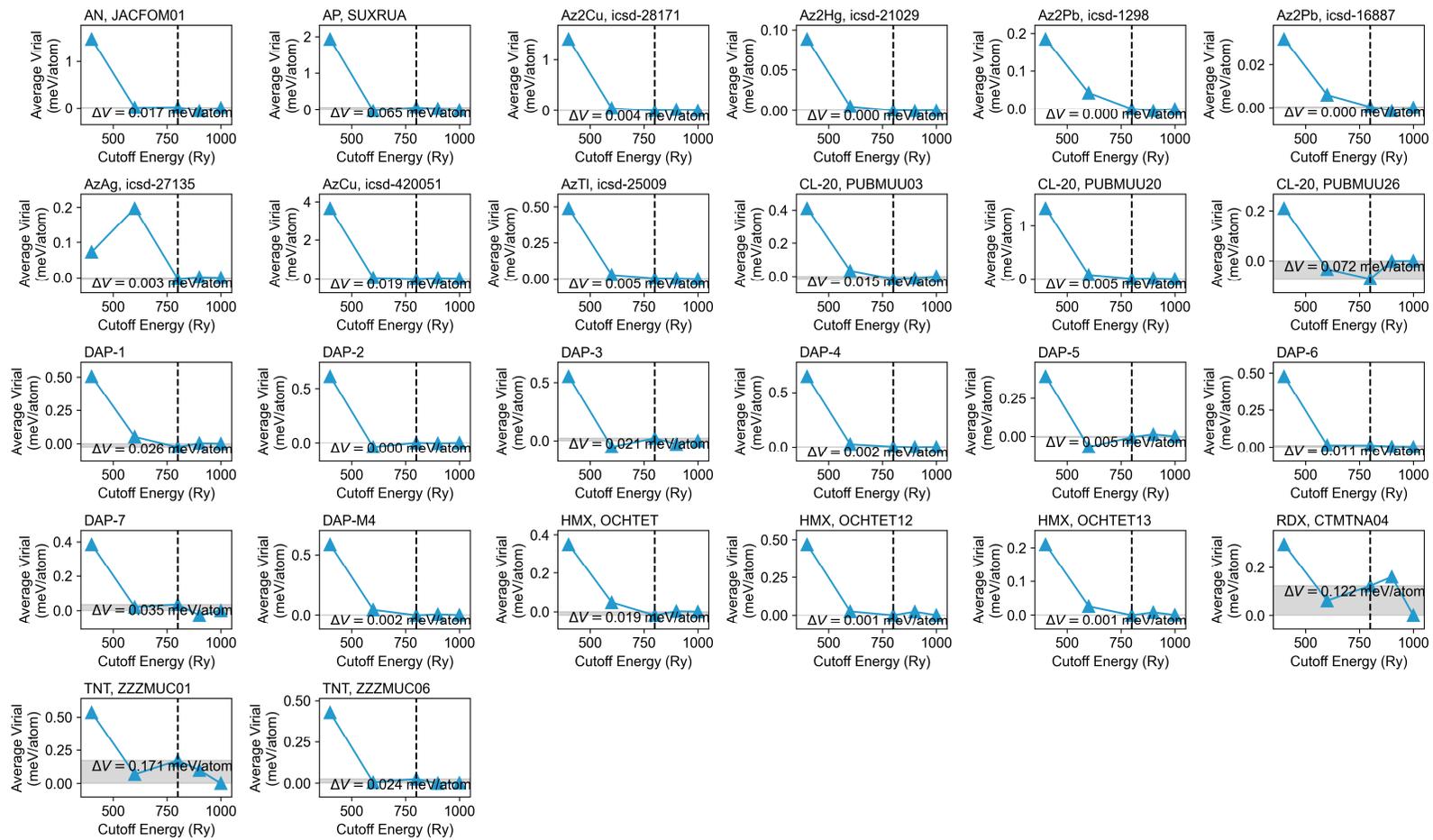

**Figure S4.** The convergence test with respect to cutoff energy of virials.

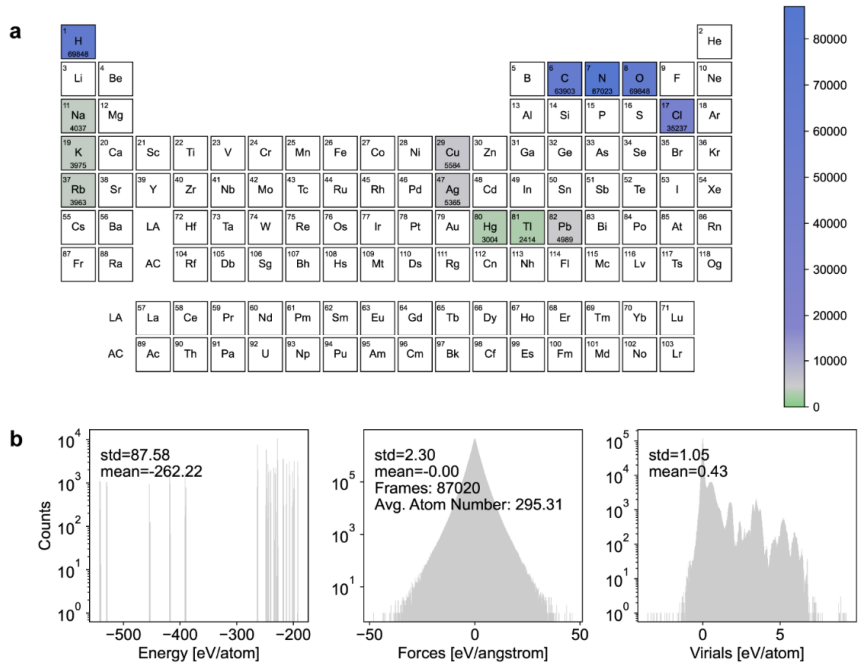

**Figure S5.** Basic information of the final datasets. (a) Element coverage and (b) DFT label distributions.

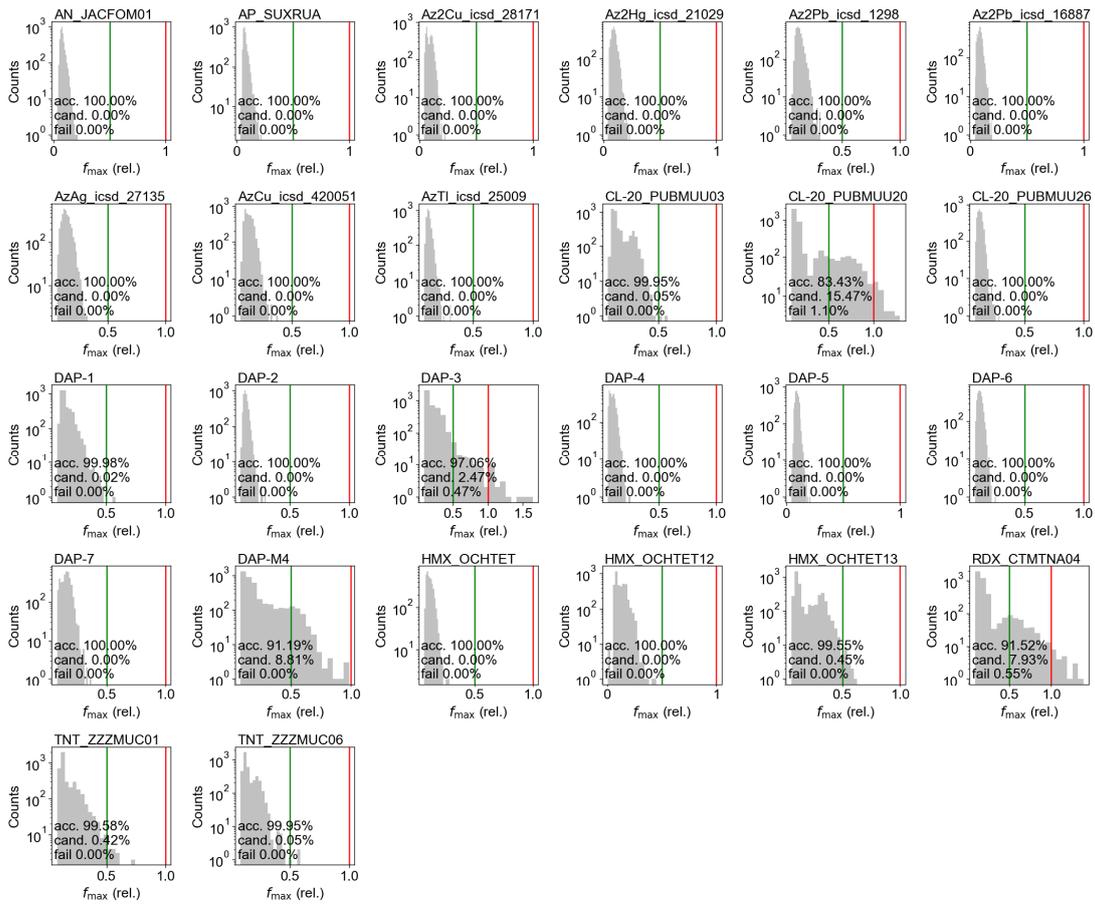

**Figure S6.** The model deviation of max forces in the last iteration of DP-GEN.

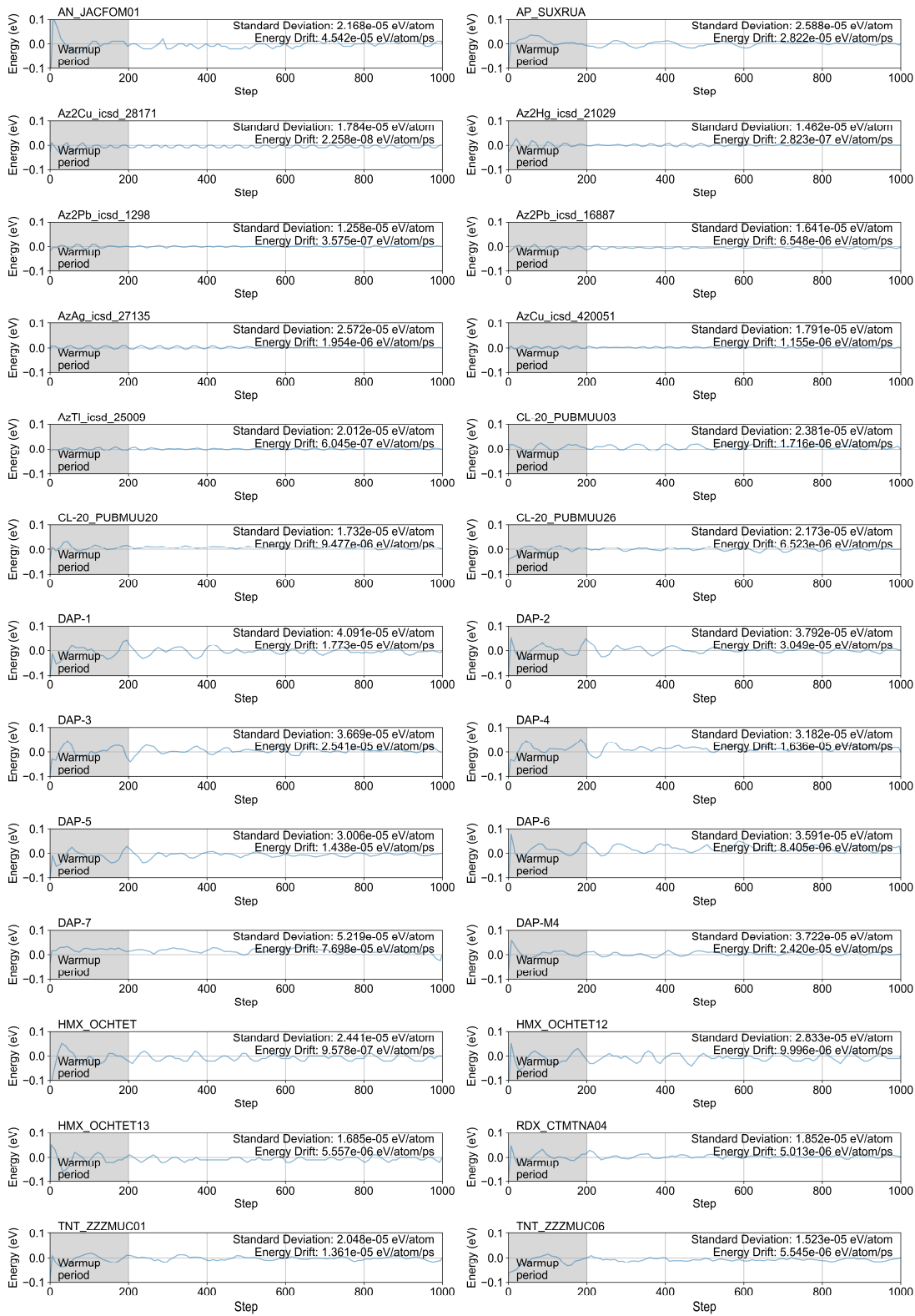

**Figure S7.** Conservativeness test. MD setting: *NVE* ensemble with timestep of 1 fs with initialized temperature of 300 K after energy minimization.

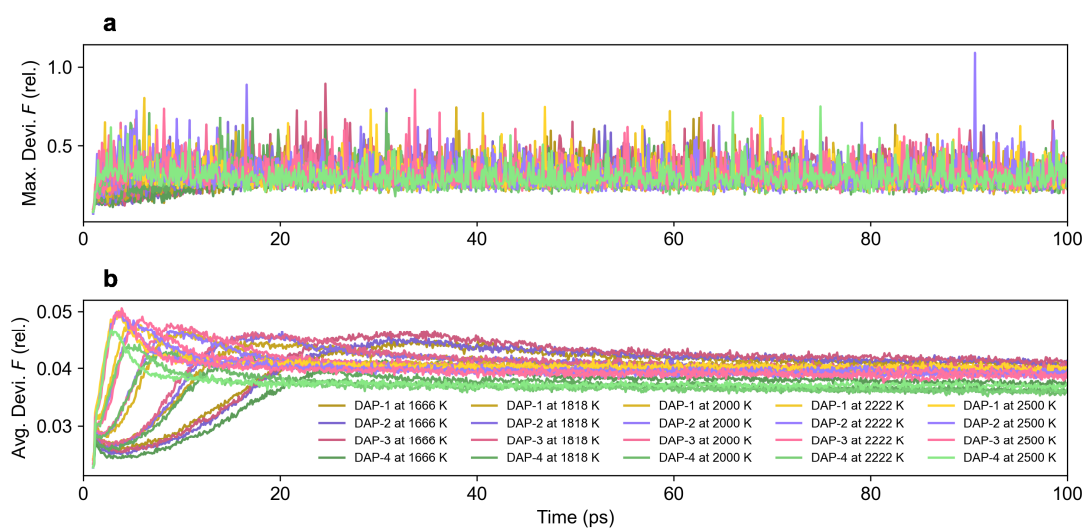

**Figure S8.** The averaged model deviations monitored during *NVT*-MD sampling.

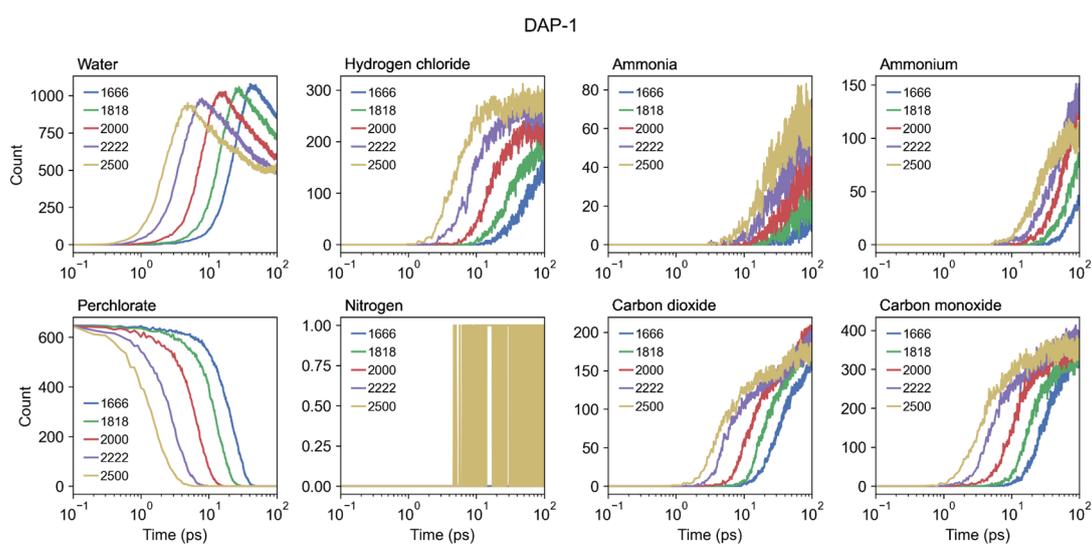

**Figure S9.** Species statistics of gaseous products from DAP-1.

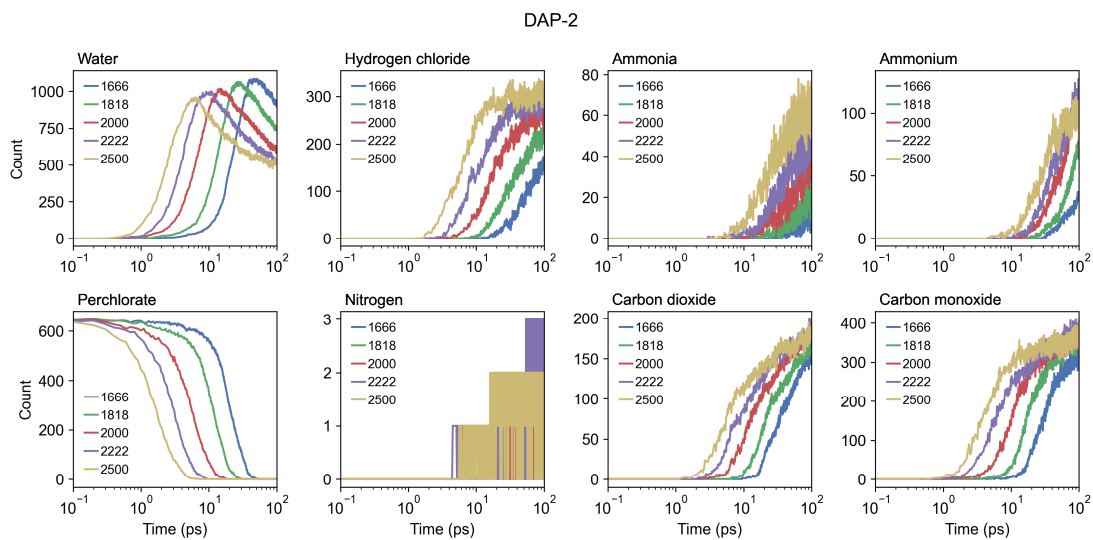

**Figure S10.** Species statistics of gaseous products from DAP-2.

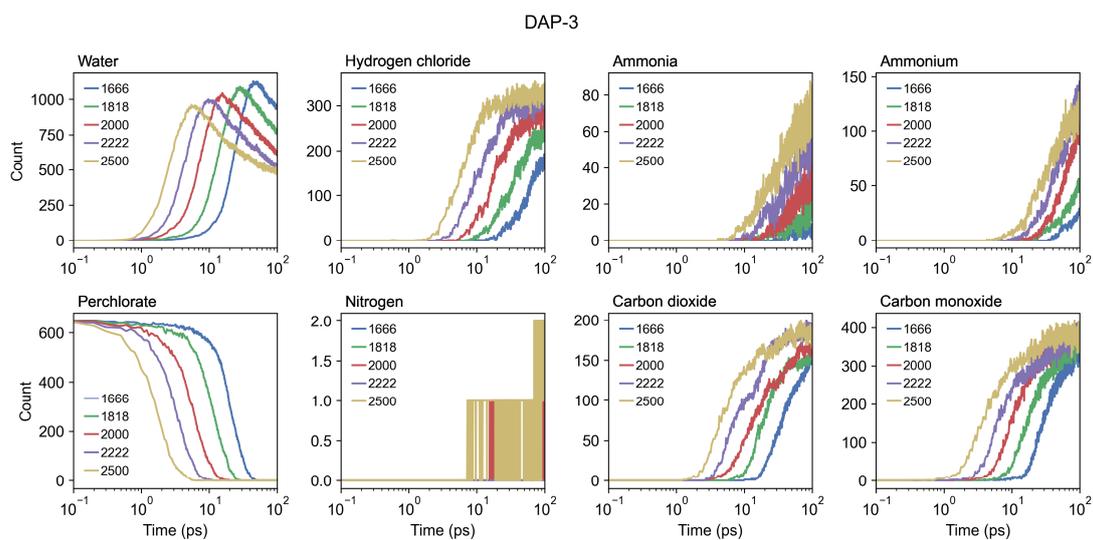

**Figure S11.** Species statistics of gaseous products from DAP-3.

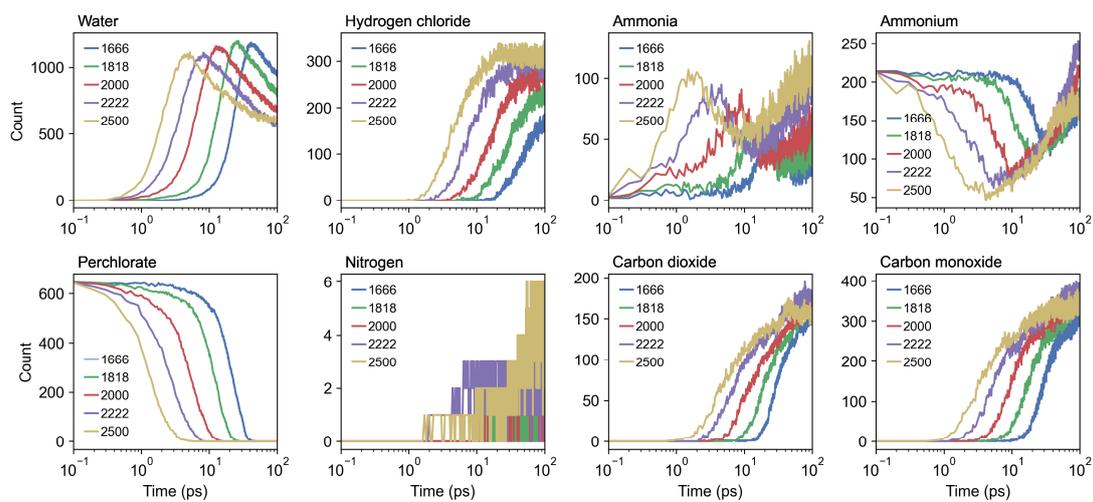

**Figure S12.** Species statistics of gaseous products from DAP-4.

# II. Supplementary Table

**Table S1.** The strategy of active-learning sampling.

| Iteration | MD Condition | MD Steps |
|---|---|---|
| 0 | *NPT-t* <br> $T$ = [100, 500]; $P$ = [0, 100000, 500000, 1200000] | 5000 |
| 1 | *NPT-t* <br> $T$ = [100, 500]; $P$ = [0, 100000, 500000, 1200000] | 10000 |
| 2 | *NVT* <br> $T$ = [100, 300, 500, 900, 1500, 2500, 3000, 4000] | 50000 |
| 3 | *NVT* <br> $T$ = [100, 300, 500, 900, 1500, 2500, 3000, 4000] | 50000 |
| 4 | *NVT* <br> $T$ = [1500, 2000, 2500, 3000, 3500, 4000] | 50000 |
| 5 | *NVT* <br> $T$ = [1500, 2000, 2500, 3000, 3500, 4000] | 50000 |
| 6 | *NVT* <br> $T$ = [1500, 2000, 2500, 3000, 3500, 4000] | 100000 |
| 7 | *NPT-i* <br> $T$ = [100, 500]; $P$ = [0, 100000, 500000, 1200000] | 100000 |
| 8 | *NPT-a* <br> $T$ = [100, 500]; $P$ = [0, 100000, 500000, 1200000] | 100000 |
| 9–17 | *NPT-t* <br> $T$ = [100, 500]; $P$ = [0, 100000, 500000, 1200000] | 100000 |
| 18–20 | *NVT* <br> $T$ = [2000, 2500, 3000, 4000] | 500000 |
| 21–22 | *NPT-t* <br> $T$ = [100, 500]; $P$ = [0, 100000, 500000, 1200000] | 100000 |

Note: *NPT-t* represents triclinic pressure control, *NPT-i* represents isotropic pressure control, *NPT-a* represents anisotropic pressure control. The units of temperature (*T*) and pressure (*P*) listed above are K and bar, respectively.

**Table S2.** The fragment recognition criteria with coordination number.

| Species | Central-Peripheral | Distance/Å | Coordination Number |
|---|---|---|---|
| Water | O–H | 1.26 | 2 |
| Hydrogen chloride | Cl–H | 1.56 | 1 |
| Ammonia | N–H | 1.27 | 3 |
| Ammonium | N–H | 1.27 | 4 |
| Perchlorate | Cl–O | 1.7 | 4 |
| Nitrogen | N–N | 1.15 | 1 |
| Carbon dioxide | C–O | 1.3 | 2 |
| Carbon monoxide | C–O | 1.2 | 1 |

**Table S3.** The fragment recognition criteria by SMILES codes.

| Species | SMILES |
|---|---|
| $H_2$dabco | [H][C]1([H])[C]([H])([H])[N]2([H])[C]([H])([H])[C]([H])([H])[N]1([H])[C]([H])([H])[C]2([H])[H] |
| $H_2$dabco-depro.C | [H][C]1[C]([H])([H])[N]2([H])[C]([H])([H])[C]([H])([H])[N]1([H])[C]([H])([H])[C]2([H])[H] |
| $H_2$dabco-depro.N | [H][C]1([H])[N]2[C]([H])([H])[C]([H])([H])[N]([H])([C]1([H])[H])[C]([H])([H])[C]2([H])[H] |
| $H_2$dabco-bridge-open | [H][C]([H])[C]([H])([H])[N]1([H])[C]([H])([H])[C]([H])([H])[N]([H])[C]([H])([H])[C]1([H])[H] |

Note: Since the bonding threshold is hard-coded in the OpenBabel library, the reactions with perchlorate here are all about the reaction of [O] with the above SMILES code to generate the corresponding dehydrogenation products and [O][H]

**Table S4.** The ionic radii of site B.

| Element | Radius/Å |
|---|---|
| $Na^+$ | 1.05 |
| $K^+$ | 1.38 |
| $Rb^+$ | 1.52 |
| $NH_4^+$ | 1.40 |